\documentclass[preprint,amsmath,amssymb,superscriptaddress,notitlepage]{revtex4-1}
\usepackage{graphicx}
\usepackage[utf8]{inputenc}
\usepackage{amssymb}
\usepackage{color}
\usepackage{units}
\usepackage{float}
\usepackage{amsmath}
\usepackage{url}
\usepackage{hyperref}
\usepackage{epsfig}
\usepackage{bm}
\usepackage{xspace} 
\hypersetup{
  colorlinks  = true, 
  urlcolor   = blue,
  linkcolor  = red, 
  citecolor  = blue 
}
\newcommand{\Rh}{TaRhTe$_4$\xspace}
\newcommand{\Ir}{TaIrTe$_4$\xspace}
\begin{document}
%%%%%%%%%%%%%%%%%%%%%%%%%%%%%%%%%%%%%%%%%%%%%%%%%%%%%%%%%%%%%%%%%%%%%%%%%%%%%%%%%%%%%%%%%%%%%%%%%%%
\title{Chiral anomaly in the  Weyl semimetal \Rh}
\author{M. Behnami$^{1,2,3}$, D. V. Efremov$^{1}$, S. Aswartham$^{1}$,  G. Shipunov$^{1}$, B. R. Piening $^{1}$, C. G. F. Blum$^{1}$, V. Kocsis$^{1}$, J. Dufouleur$^{1}$, I. Pallecchi$^{4}$, M. Putti $^{3,4}$, B. Büchner$^{1,2}$, H. Reichlova$^{1,2,5}$, and F. Caglieris$^{4}$\\ $^{1}$\textit{IFW Dresden, P.O. Box 270116, 01171 Dresden, Germany} \\$^{2}$\textit{Institut f{\"u}r Festk{\"o}rper- und Materialphysik, Technische\\ Universit{\"a}t Dresden, 01062 Dresden, Germany }\\$^{3}$\textit{Department of Physics, University of Genoa, 16146 Genova, Italy}\\$^{4}$\textit{CNR-SPIN, 16152 Genova, Italy}\\ $^{5}$\textit{Institute of Physics ASCR, v.v.i., Cukrovarnick \'a\\ 10, 162 53, Praha 6, Czech Republic} 
}
%%%%%%%%%%%%%%%%%%%%%%%%%%%%%%%%%%%%%%%%%%%%%%%%%%%%%%%%%%%%%%%%%%%%%%%%%%%%%%%%%%%%%%%%%%%%%%%%%%%
\date{\today}
%%%%%%%%%%%%%%%%%%%%%%%%%%%%%%%%%%%%%%%%%%%%%%%%%%%%%%%%%%%%%%%%%%%%%%%%%%%%%%%%%%%%%%%%%%%%%%%%%%%
\begin{abstract}
%%%%%%%%%%%%%%%%%%%%%%%%%%%%%%%%%%%%%%%%%%%%%%%%%%%%%%%%%%%%%%%%%%%%%%%%%%%%%%%%%%%%%%%%%%%%%%%%%%%
\Rh is a type-II Weyl semimetal, exhibiting four Weyl points in proximity to the Fermi level. 
In this article, we report our results of a systematic study of longitudinal magnetoresistance in \Rh. Our findings indicate that magnetoresistance becomes negative only when the magnetic field is applied parallel to the electric field.
By rotating $\mathbf{E}$ (as well as $\mathbf{B}$), we show that its origin is consistent with the prediction of the chiral anomaly, while the current jetting effect and weak localization could be excluded.
The negative magnetoresistance persists up to room temperature, suggesting that \Rh exhibits distinctive properties within the family of Weyl semimetals.
%%%%%%%%%%%%%%%%%%%%%%%%%%%%%%%%%%%%%%%%%%%%%%%%%%%%%%%%%%%%%%%%%%%%%%%%%%%%%%%%%%%%%%%%%%%%%%%%%%%
\bigskip


\end{abstract}
%%%%%%%%%%%%%%%%%%%%%%%%%%%%%%%%%%%%%%%%%%%%%%%%%%%%%%%%%%%%%%%%%%%%%%%%%%%%%%%%%%%%%%%%%%%%%%%%%%%
\maketitle
%%%%%%%%%%%%%%%%%%%%%%%%%%%%%%%%%%%%%%%%%%%%%%%%%%%%%%%%%%%%%%%%%%%%%%%%%%%%%%%%%%%%%%%%%%%%%%%%%%%
\section{Introduction}
%%%%%%%%%%%%%%%%%%%%%%%%%%%%%%%%%%%%%%%%%%%%%%%%%%%%%%%%%%%%%%%%%%%%%%%%%%%%%%%%%%%%%%%%%%%%%%%%%%%
In the last decade, tremendous attention has been devoted to the study of  Weyl and Dirac semimetals. These are novel classes of topological materials, hosting relativistic quasiparticles named Weyl and Dirac fermions \cite{yan2017topological,hasan2017discovery}. The electronic band structure of these materials is characterized by band-touching points, identified as Weyl and Dirac points, respectively, and linear quasiparticle dispersion in proximity to these points \cite{armitage2018weyl}.

In 3D Dirac semimetals (DSMs) both inversion and time reversal symmetries are present, making Dirac cones four-fold degenerate. In contrast, in Weyl semimetals (WSMs), either time-reversal symmetry (TRS) or inversion symmetry (IS) is broken, which results in Weyl points (WPs) being non-degenerate and described by a topological charge. The topological nature of WPs leads to many interesting properties, including their protection against small perturbations and distortions \cite{armitage2018weyl}. 
In addition, WSMs show many interesting phenomena, which can be observed  experimentally. One of those is the development of protected surface states, called Fermi arcs, which can be identified through angle-resolved photoemission spectroscopy and represent the fingerprint of the Weyl cones in the bulk \citep{liu2016evolution}. Another effect predicted in WSMs is the chiral anomaly, which is characterized by the anomalous non-conservation of a chiral current \cite{nielsen1983adler,yang2015chiral,huang2015observation,niemann2017chiral,li2016chiral}. 

\begin{figure}[h!tbp]
	\centering
	\includegraphics[width=\columnwidth]{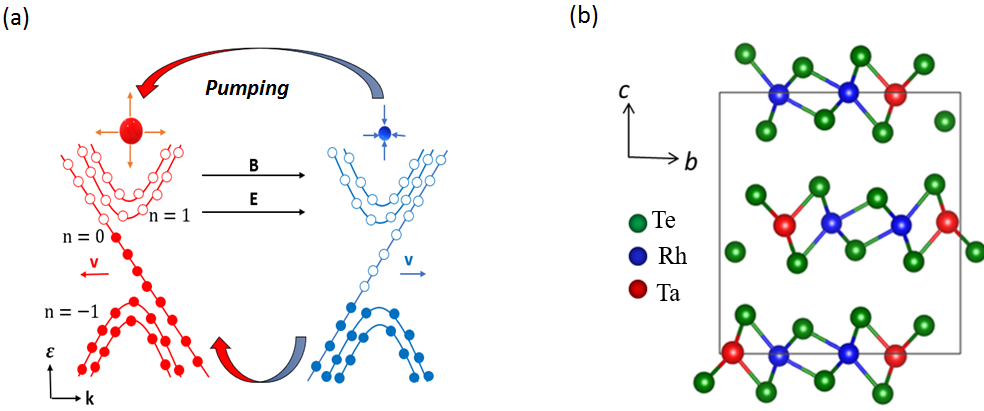}
	\caption{ (a) Schematic illustration of bulk Landau levels of a pair of Weyl
points. The linear lines represent the zeroth quantum Landau
Level with blue and red chiralities in a $\mathbf{B} \parallel \mathbf{E}$. (b) Crystal structure of the \Rh compound.}
\label{f1-1}
\end{figure}

The chiral anomaly manifests itself in transport properties as a negative contribution to the longitudinal magnetoresistance (MR) when the applied magnetic field is parallel to the electric field that causes the motion of the carriers \cite{son2013chiral,burkov2014chiral}. Figure \ref{f1-1}a shows a schematic representation of this effect. Such phenomenology has been observed in several Weyl semimetals, including NbP, NbAs \cite{Niemann2017, li2017negative}, Dirac semimetals ZrTe$_5$ \cite{li2016chiral}, Na$_3$Bi  \cite{xiong2015evidence}, Cd$_3$As$_2$ \cite{li2016negative}, GdPtBi  \cite{hirschberger2016chiral, liang2018experimental}.
In all of these cases, numerous WPs are close to the Fermi level, making it difficult to attribute the effect to any specific couple of WPs.

Here, we study \Rh, a sibling compound  of \Ir, which is the simplest example of WSMs, often referred to as the "hydrogen atom" in the world of WSMs due to its minimal number of WPs near the Fermi level, which must be four for non-centrosymmetric WSMs \cite{khim2016magnetotransport, koepernik2016tairte, belopolski2017signatures}. \Ir has many interesting properties, both from the point of view of fundamental physics and from the perspective of next-generation electronics. For instance, it shows strong positive MR at low temperatures \cite{khim2016magnetotransport}, presents room-temperature nonlinear Hall effect \cite{kumar2021room}, surface superconductivity \cite{xing2020surface} and anisotropic photo-response \cite{lai2018broadband, zhuo2021dynamical}. \Rh is much less investigated \cite{zhang2024layer}.

\Rh crystallizes in the same  orthorhombic non-centrosymmetric space group (No: Pmn2$_1$) structure as \Ir and it is a layered van der Waals (vdW) material. The crystal structure can be  derived from WTe$_2$ \cite{shipunov2021layered}, as shown in Figure \ref{f1-1}b. 
%%%%%%%%%%%%%%%%%%%%%%%%%%%%%%%%
%%%%%%%%%%%%%%%%%%%%%%%%%%%%%%%%%%%%%%%%%%%%%%%%%%%%%%%%%%%%%%%%%%%%%%%%%%%%%%%%%%%%%%%%%%%%%%%%%%%

In this article, we report our systematic investigation of the magnetotransport properties of \Rh single crystal. We measured the MR signal by applying a magnetic field up to 10 T in various geometries, including in-plane (parallel and perpendicular to the electric field) and out-of-plane (perpendicular to the electric field). Surprisingly, the magnetotransport properties of \Rh strongly differ from \Ir (shown in supplementary information Fig. S1 \cite{SIMahdi}) despite the similarity of the crystal structures and electronic band structures \cite{shipunov2021layered}.   
In particular, in the \Rh sample, we observe a negative MR when the electric and magnetic fields are parallel to each other, either oriented along the crystallographic a- and b- directions. Furthermore, this negative MR persists up to room temperature, suggesting that \Rh
possesses unique properties within the family of WSMs.
\section{Experimental}
%%%%%%%%%%%%%%%%%%%%%%%%%%%%%%%%%%%%%%%%%%%%%%%%%%%%%%%%%%%%%%%%%%%%%%%%%%%%%%%%%%%%%%%%%%%%%%%%%%%
The single crystals were grown using the self-flux method. To prepare the crystals for further study, the surfaces, which were contaminated by a small amount of flux, were mechanically cleaved \cite{shipunov2021layered}. The orientation of the crystal axis has been systematically determined by x-ray Laue diffraction. \\
The setup for DC electric transport measurements was prepared by gluing the electrodes (silver wires with 50 $\mu m$ thick) to the samples using silver paint (Dupont 4929N) as shown in the inset Figure \ref{f2-1}a. 
All transport measurements were carried out within a temperature range from 8~K to room temperature, using a tested homemade probe and Oxford cryostats equipped with sufficiently large magnets.
A Keithley Standard Series 2400 Source Measure Unit (SMU) was used to apply the electric current, while a Keithley Nanovoltmeter Model 2182A was utilized to measure the voltage signal. \textcolor{red}{A semi-sum of the datapoints acquired with positive and negative electric currents has been systematically performed in all the measurements to eliminate the thermoelectric background.}\\
The magnetoresistance has been measured in magnetic field up to 10~T, applied both in-plane (parallel and perpendicular to the electric field) and out-of-plane (perpendicular to the electric field). 
%%%%%%%%%%%%%%%%%%%%%%%%%%%%%%%%%%%%%%%%%%%%%%%%%%%%%%%%%%%%%%%%%%%
\section{Results}
\begin{figure}[h!tbp]
	\centering
	\includegraphics[width=\columnwidth]{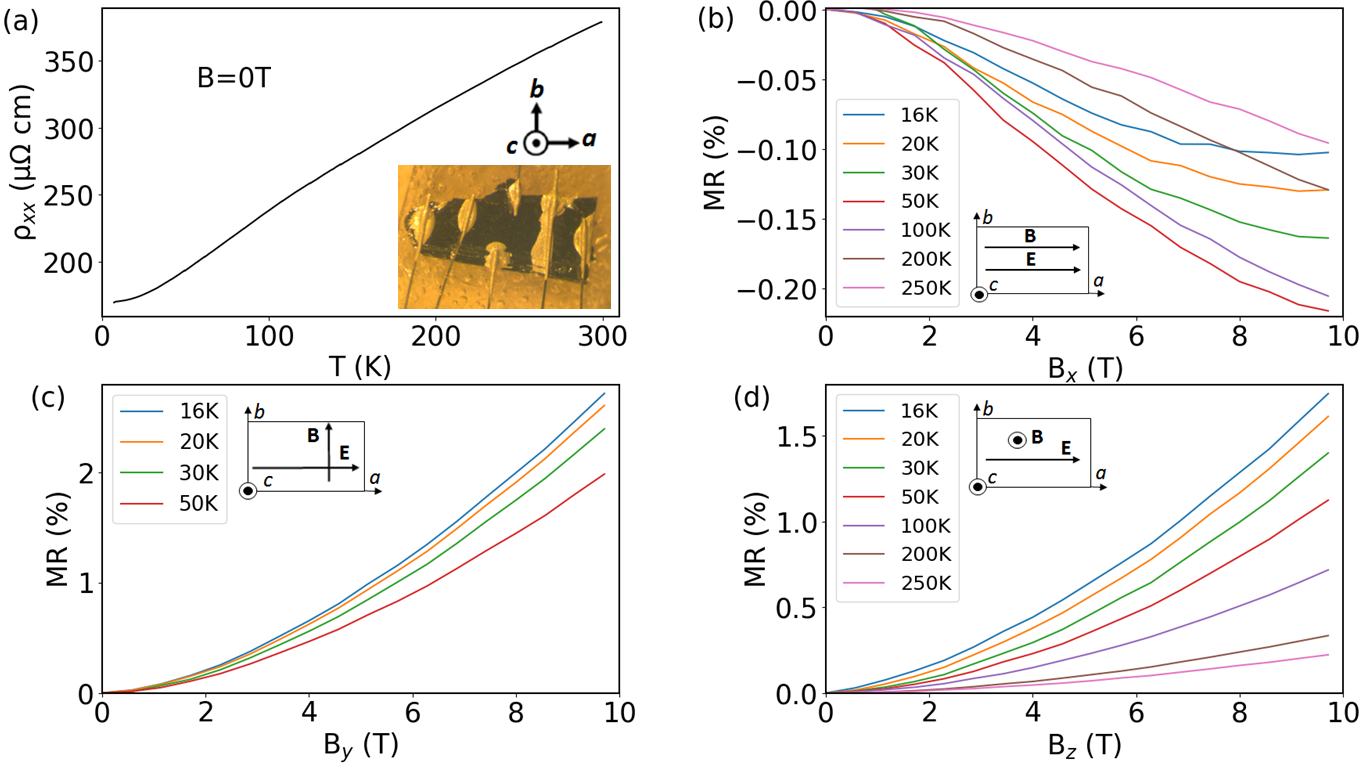}
	\caption{Longitudinal magnetoresistance (MR) in the first sample of \Rh. (a) Temperature dependence of resistivity. Inset:  a microscopic image of our  first \Rh single crystal was prepared for electric measurement. (b) Longitudinal MR in the case of an in-plane magnetic field with $\mathbf{B}\parallel \mathbf{E}$. (c)  Longitudinal MR in the case of an in-plane magnetic field with $\mathbf{B}\perp \mathbf{E}$. (d) Longitudinal MR in the case of out-of-plane magnetic field with $\mathbf{B}\perp \mathbf{E}$.}
\label{f2-1}
\end{figure}

%%%%%%%%%%%%%%%%%%%%%%%%%%%%%%%%%%%%%%%%%%%%%%%%%%%%%%%%%%%%%%%%%%%%%%%%%%%%%%%%%%%%%%%%%%%%%%%%%%%
Figure~\ref{f2-1} shows the magnetotransport measurements for the first sample of \Rh. 
More in detail, Figure~\ref{f2-1}a presents the temperature dependence of the electrical resistivity in a zero magnetic field. The curve shows a metallic behavior decreasing monotonically with the temperature and a moderate RRR= 2.23.
Figures~\ref{f2-1}b, c, and d show the dependence of resistivity on the magnetic field. 
In this experiment, the electric current was injected along the crystallographic $a$-axis and, as a consequence, the electric field results oriented along the same axis. The magnetic field was applied sequentially along all three crystallographic axis, as shown in the inset of Figure~\ref{f2-1}a. The presented MR is defined as MR $=100\times \frac{\rho \,(\mathbf{B})- \rho\, (0)}{\rho \,(0)}$, where $\rho \,(\mathbf{\mathbf{B}})$ and $\rho \,(0)$ are the resistivities of the sample in a magnetic field $\mathbf{B}$ and in zero magnetic field, respectively.

Figure~\ref{f2-1}b shows the resistivity as a function of the magnetic field ($\mathbf{B}$), where $\mathbf{B}$ is in ab-plane and parallel to $\mathbf{E}$. The MR is negative over the whole range of studied temperatures. However, at $T$= 200~K and 250~K, the curves exhibit a parabolic shape up to 10 T, while for $T\le$ 100 K, MR$\sim \mathbf{B}^2$ only in the low-field region and a change of concavity with a tendency to an upturn occurs at high fields. At $T=50$ K and $\mathbf{B}=10$ T, a negative MR of 0.21 $\%$ is obtained. The measured behavior is very similar to that observed in NbP \cite{niemann2017chiral} and Cd$_3$As$_2$ \cite{li2016negative}. Interestingly, the sibling compound \Ir does not present any negative magnetoresistance, as shown in supplementary information Fig. S1 \cite{SIMahdi}.

Figures \ref{f2-1}c and \ref{f2-1}d present the magnetoresistance as a function of the magnetic field applied along the crystal $b$- and $c$-axis, respectively. In both cases, MR is always positive and relatively small, reaching the maximum value of a few percent. In addition, it presents a parabolic behavior, which is indicative of the cyclotronic origin of the magnetoresistive effect \cite{li2016negative}. 
%%%%%%%%%%%%%%%%%%%%%%%%%%%%%%%%%%%%%%%%%%%%%%%%%%%%%%%%%%%%%%%%%%%%%%%%%%%%%%%%%%%%%%%%%%%%%%%%%%%
%%%%%%%%%%%%%%%%%%%%%%%%%%%%%%%%%%%%%%%%%%%%%%%%%%%%%%%%%%%%%%%%%%%%%%%%%%%%%%%%%%%%%%%%%%%%%%%%%%%

\begin{figure}[h!tbp]
	\centering
	\includegraphics[width=0.55\columnwidth]{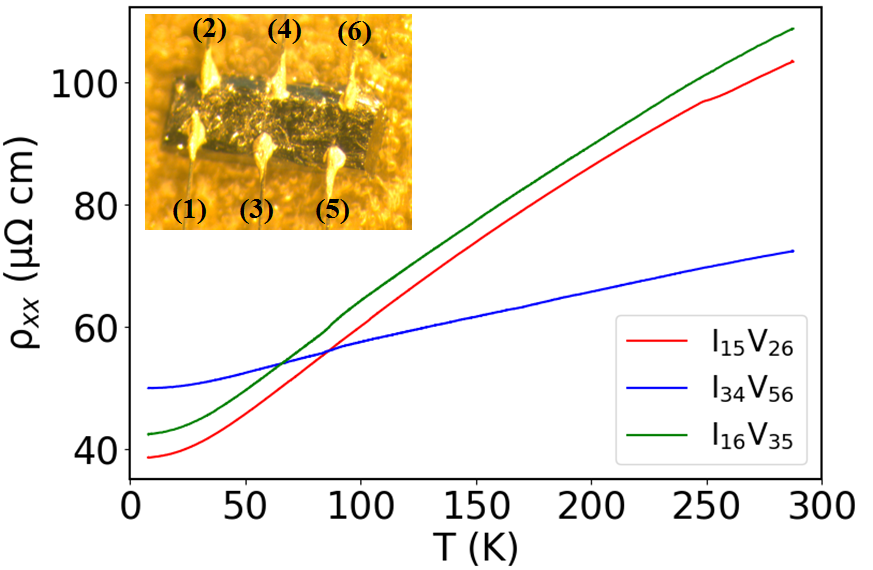}
	\caption{Temperature dependence of resistivity for three configurations of the second sample of \Rh. Inset: A microscopic picture of the second sample of \Rh. The numbers indicate the electrodes that we used to apply the electrical field and measure longitudinal resistivity.}
\label{f6-1}
\end{figure}

In order to confirm the negative MR, observed in Figure ~\ref{f2-1}b, we deeply investigated a second sample. %The second sample of \Rh has shown a negative MR as well.
To this aim, we chose a slab-shaped sample of \Rh and we attached six electrodes on top of that, which enables us to measure the MR signal along different crystallography axis (see inset of Figure \ref{f6-1}). We first measured the zero-field resistivity of the second sample in different configurations, as shown in Figure \ref{f6-1}. Interestingly, the resistivity estimated with voltage electrodes along the $a$-axis (V$_{26}$ and V$_{35}$) and current prevalently injected along the $a$-axis is relatively different with respect to the resistivity evaluated putting both voltage and current electrodes along the sample $b$-axis (I$_{34}$ and V$_{56}$). 
The different RRR values (2.68, 2.55, and 1.44) indicate the relative  anisotropy between the $a$-axis and the $b$-axis conductivities. From the finite element simulations (see supplementary information Fig. S2) and from the comparison of the experimental values of zero-field resistances measured in the three configurations, we extracted the value of in-plane resistivity anisotropy $\rho_{b}/\rho_{a}$=6.8.

\begin{figure}[h!tbp]
	\centering
	\includegraphics[width=\columnwidth]{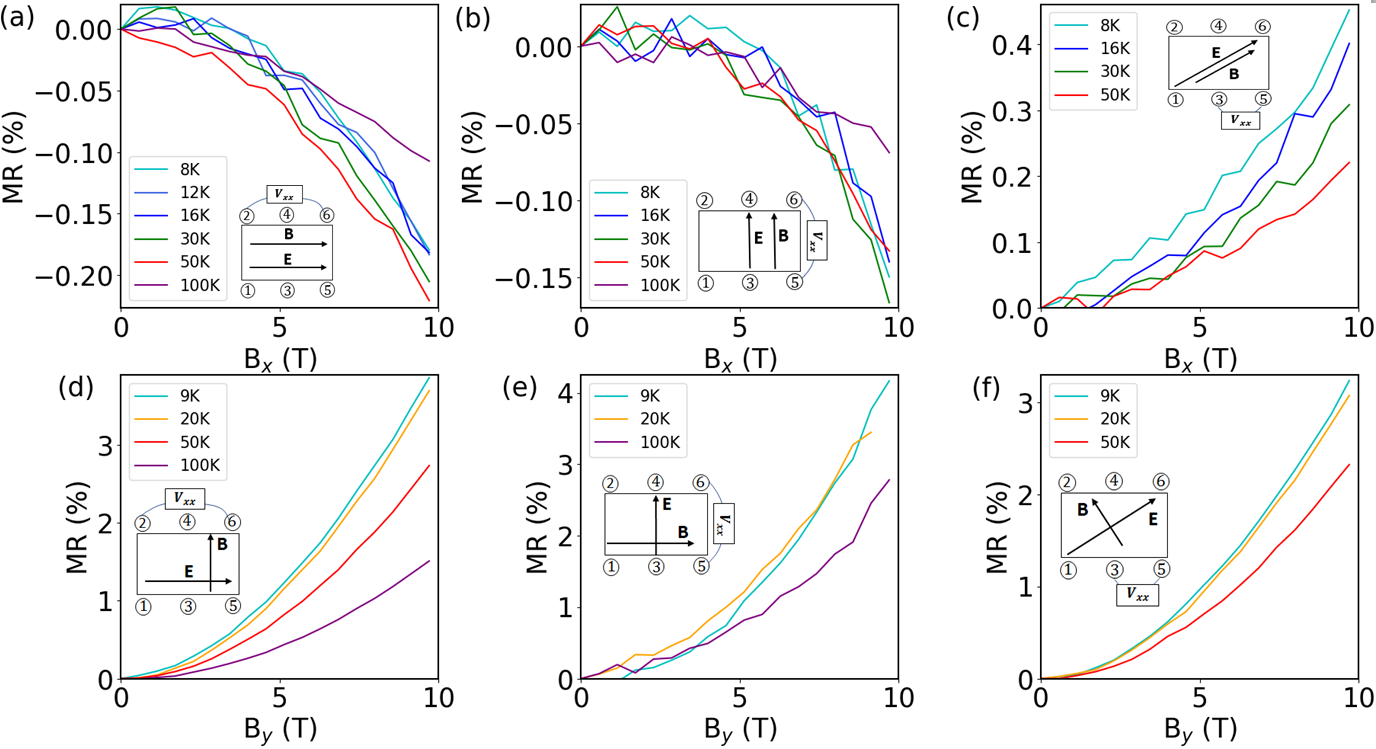}
	\caption{Longitudinal MR in the second sample of \Rh. 
(a-c) Longitudinal MR in the case of an in-plane magnetic field with $\mathbf{B}\parallel \mathbf{E}$. The applied magnetic field is along $a$-, $b$-, and  30 degrees from the $a$-axis of the crystal respectively. (d-f)  Longitudinal MR in the case of an in-plane magnetic field with $\mathbf{B}\perp \mathbf{E}$. The applied magnetic field is along $b$-, $a$-, and  30 degrees from the $b$-axis of the crystal respectively. }
\label{f3-1}
\end{figure}

After a zero-field characterization, we measured the sample magnetoresistance. For this specimen, we explored the following configurations.
%%%%%%%%%%%%%%%%%%%%%%%%%%%%%%%%%%%%%%%%%%%%%%%%%%%%%%%%%%%%%%%%%%%%%%%%%%%%%%%%%%%%%%%%%%%%%%%%%%%
\paragraph{ Magnetic field in the ab-plane and electric field parallel to the $a$-axis of the crystal.}
%%%%%%%%%%%%%%%%%%%%%%%%%%%%%%%%%%%%%%%%%%%%%%%%%%%%%%%%%%%%%%%%%%%%%%%%%%%%%%%%%%%%%%%%%%%%%%%%%%%
In the first setup, we injected current from contact (1) to contact (5) and measured the voltage signal between contact (2) and contact (6), with the magnetic field applied parallel to the $a$-axis of the crystal. As illustrated in Figure \ref{f3-1}a, we observed a negative MR, which reaches the maximum value of 0.22 \% and persists up to high temperatures. In the second setup, we changed the direction of the magnetic field from parallel to perpendicular to the $a$-axis. This time, we observed a positive MR which reaches 3.9 \% for $T=9$ K and $\mathbf{B}=10$ T, as depicted in Figure \ref{f3-1}d. 
%%%%%%%%%%%%%%%%%%%%%%%%%%%%%%%%%%%%%%%%%%%%%%%%%%%%%%%%%%%%%%%%%%%%%%%%%%%%%%%%%%%%%%%%%%%%%%%%%%%
 \paragraph{Magnetic field in the ab-plane and electric field parallel to the $b$-axis of the crystal.}
 %%%%%%%%%%%%%%%%%%%%%%%%%%%%%%%%%%%%%%%%%%%%%%%%%%%%%%%%%%%%%%%%%%%%%%%%%%%%%%%%%%%%%%%%%%%%%%%%%%
 In this configuration, we injected current from contact (3) to contact (4) and measured the voltage signal between contact (5) and contact (6). Firstly, we applied a magnetic field parallel to the $b$-axis of the crystal, observing again a negative MR (up to 0.19 \%) as illustrated in Figure \ref{f3-1}b. Secondly, we changed the direction of the magnetic field from parallel to perpendicular to the sample $b$-axis. This time, we observed a positive MR up to 4.2 \% at $T=9$ K and $\mathbf{B}=10$ T, as depicted in Figure \ref{f3-1}e. 
%%%%%%%%%%%%%%%%%%%%%%%%%%%%%%%%%%%%%%%%%%%%%%%%%%%%%%%%%%%%%%%%%%%%%%%%%%%%%%%%%%%%%%%%%%%%%%%%%%%
\paragraph{Magnetic field in the ab-plane and electric field at 30 degrees from $a$-axis of the crystal:}
%%%%%%%%%%%%%%%%%%%%%%%%%%%%%%%%%%%%%%%%%%%%%%%%%%%%%%%%%%%%%%%%%%%%%%%%%%%%%%%%%%%%%%%%%%%%%%%%%%%
In the third configuration, we injected current from contact (1) to contact (6) and measured the voltage signal between contact (3) and contact (5). In the first experiment, we applied a magnetic field parallel to both an electric field and the diagonal axis of the sample, tilted about 30 degrees from the $a$-axis. Clearly, such an angle is only a nominal value, not necessarily corresponding to the exact direction of the electric field, which strictly depends on the actual current path. As illustrated in Figure \ref{f3-1}c, we observed a positive MR. At $T=8$ K and $\mathbf{B}=10$ T, the highest positive MR of 0.45 \% can be obtained. In the second configuration, we changed the direction of the magnetic field from parallel to perpendicular and measured the signal. Once again, we observed a positive MR, as depicted in Figure \ref{f3-1}f. At $T=9$ K and $\mathbf{B}=10$ T, a positive MR of 3.2 \% is measured. Therefore, we did not observe negative MR in this setup, whether the magnetic field was parallel or perpendicular. 
%%%%%%%%%%%%%%%%%%%%%%%%%%%%%%%%%%%%%%%%%%%%%%%%%%%%%%%%%%%%%
%%%%%%%%%%%%%%%%%%%%%%%%%%%%%%%%%%%%%%%%%%%%%%%%%%%%%%%%%%%%%%%%%%%%%%%%%%%%%%%%%%%%%%%%%%%%%%%%%%%
\section{Discussion}
%%%%%%%%%%%%%%%%%%%%%%%%%%%%%%%%%%%%%%%%%%%%%%%%%%%%%%%%%%%%%%%%%%%%%%%%%%%%%%%%%%%%%%%%%%%%%%%%%%%
In non-magnetic conductors, the presence of external magnetic fields typically generates a magnetoresistive effect caused by the cyclotronic motion of the charge carriers, when the field itself has a component perpendicular to the direction of the current density vector \cite{li2016negative}. Such semiclassical effect strongly depends on both intrinsic and extrinsic properties of the measured sample, such as the effective mass of the charge carriers and the scattering time, and, in the low field limit, is proportional to B$^2$. In general, clean systems exhibit higher cyclotronic magnetoresistance, intuitively depending on the number of cyclotronic orbits traveled by charge carriers between a scattering event and the following one. As expected, in our case, this is also the dominant contribution in all the configurations with the magnetic and electric fields not aligned in the same direction.

\begin{figure}[h!tbp]
	\centering
	\includegraphics[width=\columnwidth]{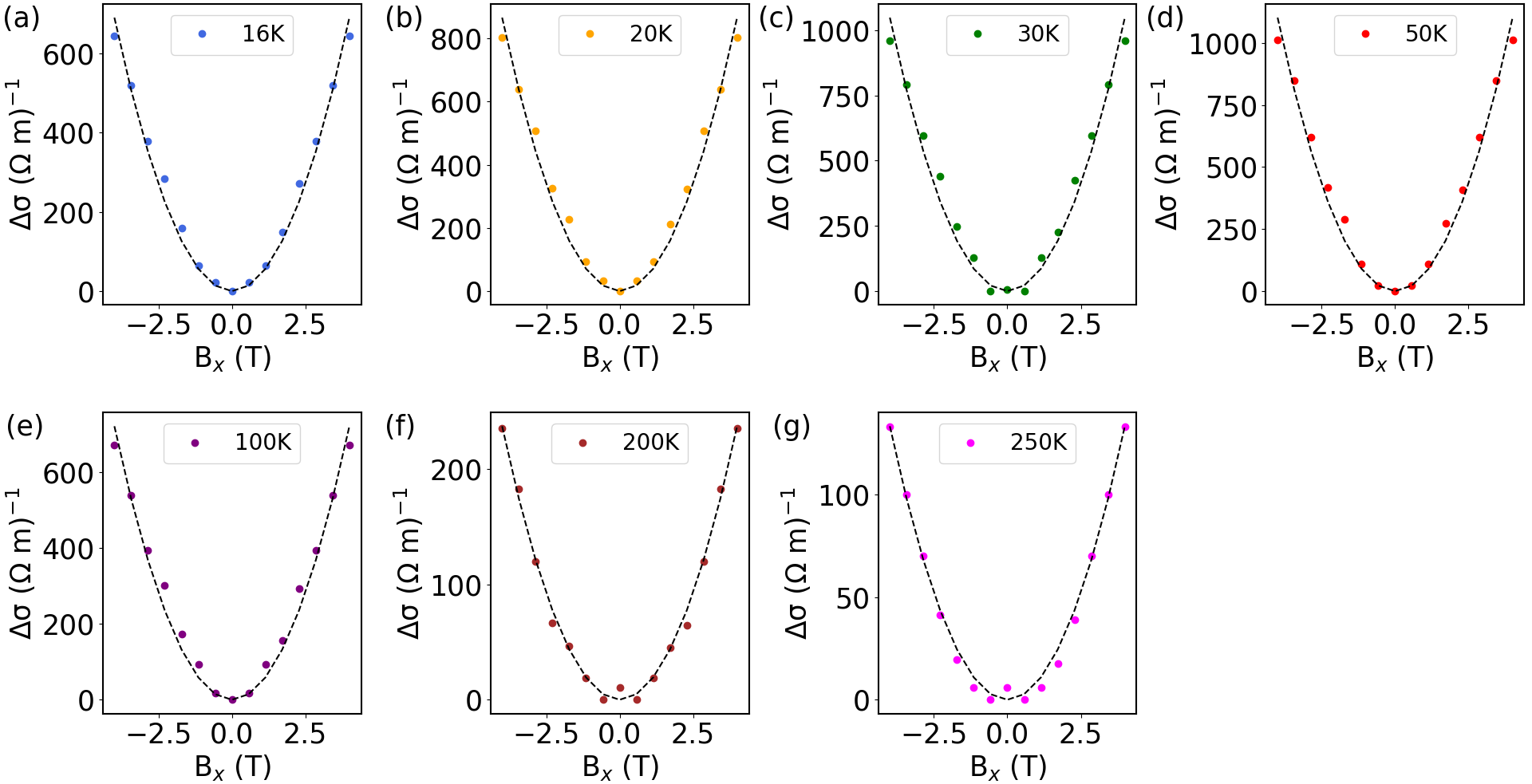}
	\caption{Positive contribution of electrical conductivity as a function of  magnetic field for the first sample of \Rh. The direction of a magnetic field is parallel to the electric field ($\mathbf{B} \parallel \mathbf{E}$).  }
\label{f4-1}
\end{figure}

In Weyl semimetals, an additional negative contribution to the magnetoresistance is possibly caused by the chiral anomaly. Equivalently, the chiral anomaly manifests itself as a positive contribution to the electrical conductivity in magnetic field \cite{son2013chiral, burkov2016topological}, only when it is parallel to the electric field that causes the motion of charge carriers, namely to the direction of the injected current. In this situation, the correction to the electrical conductivity due to the chiral anomaly for a type-I Weyl semimetal reads \cite{son2013chiral}:  
%%%%%%%%%%%%%%%%%%%%%%%%%%%%%%%%%%%%%%%%%%%%%%%%%%%%%%%%%%%%%%%%%%%%%%%%%%%%%%%%%%%%%%%%%%%%%%%%%%%
\begin{equation}
	\Delta \sigma (B)  = \frac{e^2 }{4 \pi^2 \hbar c} \frac{v_F}{c} \frac{(eBv_F)^2 }{\epsilon_F^2} \tau_a,
\end{equation}
%%%%%%%%%%%%%%%%%%%%%%%%%%%%%%%%%%%%%%%%%%%%%%%%%%%%%%%%%%%%%%%%%%%%%%%%%%%%%%%%%%%%%%%%%%%%%%%%%%%
where $\tau^{-1}_a$ is the longitudinal axial current relaxation rate, $c$ is the light speed and $\epsilon_F$ and $v_F$ are the Fermi energy and the Fermi velocity respectively. Remarkably, it has been demonstrated that the quadratic-in-field dependence holds also 
in case of a type-II, or tilted WSMs, for which a positive longitudinal magneto-conductivity appears for all finite angles between the tilt direction of the cones and the direction of the applied magnetic/electric field \cite{sharma2017chiral}.
 Figure \ref{f4-1} presents the positive magnetoconductivity of the first sample of \Rh in the experimental configuration of Figure~\ref{f2-1}b. All the curves are nicely reproduced by the $\mathbf{B}^2$ fitting function, corroborating the chiral anomaly as a possible source for the observed effect. Here,  $\Delta \sigma = \sigma(\mathbf{B}) -\sigma(\mathbf{B}=0)$, where $\sigma(\mathbf{B})$ and $\sigma(\mathbf{B=0})$ are the conductivity of the first sample in a magnetic field and in zero magnetic fields respectively. The same result is obtained for the second sample as shown in supplementary information Figures S3 and  S4 \cite{SIMahdi}.

In the following, we will briefly discuss other possible sources of negative magnetoresistance in non-magnetic materials.
The first possible phenomenon that can cause negative MR is the weak localization effect \cite{altshuler1980magnetoresistance, lee1985disordered,qu2024observation}.However, weak localization is typically observed in 2D or highly anisotropic (out-of-plane/in-plane anisotropy of electronic properties) materials, when the magnetic field is applied in the out-of-plane direction. In this condition, the threading of magnetic flux in interfering current paths may result in negative magnetoresistance. In our case, the negative MR appears when field and current are both in-plane and parallel to each other, which is the configuration that minimizes the threading of magnetic flux in interfering current paths. Moreover, both the magnitude and the magnetic field dependence of our magnetoresistance are not described by conventional models for weak localization (see Supplemental Information\cite{SIMahdi}).

%Moreover weak localization should be observed for any direction of the magnetic field, including when the magnetic field is perpendicular to the current direction, a configuration in which our magnetoresistance is strictly positive.
The main alternative to the chiral anomaly to explain the observed phenomenology is represented by the so-called current jetting effect, which appears as a semiclassical negative longitudinal MR caused by a non-uniform current distribution in the sample when $\mathbf{B} \parallel \mathbf{E}$. Such an effect typically appears when current contacts are smaller than the cross-section of the sample, causing field-induced steering of the current to the direction of the magnetic field \cite{liang2018experimental, sykora2020disorder}. In the measurement configuration of Figure~\ref{f2-1}b, referring to sample one, the size of the current electrodes is comparable with the sample width (see inset of Figure~\ref{f2-1}a). Therefore, we do not expect any substantial current jetting. Remarkably, in the case of sample two, where the electrodes are smaller and differently distributed across the sample (see inset of Figure \ref{f6-1}), the experimental configurations in Figure \ref{f3-1}a and \ref{f3-1}b returns almost the same value of negative magnetoresistance, very similar to the one observed in sample one. In addition, we tested the influence of size and position of the electrical contacts on the appearance of the negative MR by measuring several other samples. In all the experiments we verified that the only relevant parameter is the mutual orientation of current, magnetic field and crystal axis, which have to be parallel to each other, as expected in case of chiral anomaly (see Supplementary Information\cite{SIMahdi}).\\
Finally, we note that, in sample one, at $T<$100 K the negative magnetoresistive curves of Figure~\ref{f2-1}b change their concavity, with a tendency to turn up at high fields. This effect can be trivially explained through an involuntary tilting of the applied magnetic field towards the out-of-plane direction, which causes an additional positive cyclotronic component that becomes dominant for sufficiently large fields. This spurious effect also accounts for the non monotonic trend in temperature of the curves in Figure 2b. The curves a $T$=200 and 250 K do not show the upturn due to the strong suppression of the cyclotronic magnetoresitance at high temperatures.
Hence, the insensitivity of the observed magnetoresistance to the geometry and distribution of the current electrodes on the sample excludes with reasonable certainty the current jetting as a possible source for the negative magnetoresistance, in favor of an intrinsic phenomenon related to its non-trivial topology.
%%%%%%%%%%%%%%%%%%%%%%%%%%%%%%%%%%%%%%%%%%%%%%%%%%%%%%%%%%%%%%%%%%%%%%%%%%%%%%%%%%%%%%%%%%%%%%%%%%%
\section{Conclusions}
%%%%%%%%%%%%%%%%%%%%%%%%%%%%%%%%%%%%%%%%%%%%%%%%%%%%%%%%%%%%%%%%%%%%%%%%%%%%%%%%%%%%%%%%%%%%%%%%%%%
In conclusion, we have systematically measured the MR of two single crystals of \Rh, with different experimental configurations by varying both the distribution of the current/voltage electrodes across the samples and the applied magnetic field direction.
In the setups where the applied electric and magnetic fields are parallel to each other and aligned either along the $a$- or the $b$- axis, we always measured a negative magnetoresistance (or positive magnetoconductance) compatible with a chiral anomaly effect. The robustness of the observed phenomenology with respect to the change of geometry in the current electrodes allowed us to exclude the current jetting as an alternative source for the effect.
%%%%%%%%%%%%%%%%%%%%%%%%%%%%%%%%%%%%%%%%%%%%%%%%%%%%%%%%%%%%%%%%%%%%%%%%%%%%%%%%%%%%%%%%%%%%%%%%%%%

\medskip
\textbf{Supporting Information} \par 
Supporting Information is available from the Online Library or from the author \cite{SIMahdi}.

\medskip
\textbf{Acknowledgements} \par 
%%%%%%%%%%%%%%%%%%%%%%%%%%%%%%%%%%%%%%%%%%%%%%%%%%%%%%%%%%%%%%%%%%%%%%%%%%%%%%%%%%%%%%%%%%%%%%%%%%%
We would like to thank Tino Schreiner and Danny Baumann for their technical support. 
MB acknowledges Forschungsstipendien für Doktorandinnen und Doktoranden, 2024 (DAAD) via project number 57694190 and the IFW Excellence Program 2018. SA acknowledges DFG via project number 523/4-1. We would like to acknowledge the financial support provided by the Deutsche Forschungsgemeinschaft (DFG) through Project number 449494427. This work was supported by the
Deutsche Forschungsgemeinschaft (DFG, German
Research Foundation) under Germany’s Excellence Strategy
through the Wurzburg-Dresden Cluster of Excellence
on Complexity and Topology in Quantum Matter—ct.qmat
(EXC 2147, project ID 242021). The project was supported by the Leibniz Association through the Leibniz Competition.
%%%%%%%%%%%%%%%%%%%%%%%%%%%%%%%%%%%%%%%%%%%%%%%%%%%%%%%%%%%%%%%%%%%%%%%%%%%%%%%%%%%%%%%%%%%%%%%%%%%
\medskip
\textbf{Data Availability}\par
The data that support the findings of this article are openly available \cite{DataAvailability}

\medskip
\textbf{Conﬂict of Interest}\par
The authors declare no conﬂict of interest.

\medskip
\textbf{Author Contributions}\par
%%%%%%%%%%%%%%%%%%%%%%%%%%%%%%%%%%%%%%%%%%%%%%%%%%%%%%%%%%%%%%%%%%%%%%%%%%%%%%%%%%%%%%%%%%%%%%%%%%%
B.B., D.V.E., H.R., F.C., and M.B. proposed the study. G.Sh., B.R.P., C.G.F.B. and S.A. prepared and characterized the samples. M.B., H.R., V.K., M.P., J.D., I.P., D.V.E., and F.C. designed the measurement setup, performed the experiments, and analyzed the data. I.P. performed the finite element simulations. B.B., H.R., and F.C. supervised the study. M.B., H.R., D.V.E., and F.C.  wrote the manuscript with input from all authors.
%%%%%%%%%%%%%%%%%%%%%%%%%%%%%%%%%%%%%%%%%%%%%%%%%%%%%%%%%%%%%%%%%%%%%%%%%%%%%%%%%%%%%%%%%%%%%%%%%%%

\medskip
% \bibliographystyle{unsrtnat}
% \bibliography{myref}
%%%%%%%%%%%%%%%%%%%%%%%%%%%%%%%%%%%%%%%%%%%%%%%%%%%%%%%%%%%%%%%%%%%%%%%%%%%%%%%%%%%%%%%%%%%%%%%%%%%

\end{document}